\documentclass[conference]{IEEEtran}
\IEEEoverridecommandlockouts
\usepackage{cite}
\usepackage{amsmath,amssymb,amsfonts}
\usepackage{algorithmic}
\usepackage{graphicx}
\usepackage{textcomp}
\usepackage{xcolor}
\usepackage{longtable}
\usepackage{booktabs}
\usepackage{graphicx}
\usepackage{tabularx}
\usepackage{multirow}
\usepackage{array}
\usepackage{float}
\usepackage{stfloats} 
\usepackage{hyperref}
\usepackage{listings}
\usepackage{tabularx}
\usepackage{caption}

\usepackage[linesnumbered,ruled,vlined]{algorithm2e}
\usepackage{hyperref}
\usepackage{cleveref}

\input{macros}

\def\BibTeX{{\rm B\kern-.05em{\sc i\kern-.025em b}\kern-.08em
    T\kern-.1667em\lower.7ex\hbox{E}\kern-.125emX}}
\crefname{algorithm}{Algorithm}{Algorithms}
\Crefname{algorithm}{Algorithm}{Algorithms}

\begin{document}

\title{ACER: An AST-based Call Graph Generator Framework\\
}

\author{
  \IEEEauthorblockN{Andrew Chen}
  \IEEEauthorblockA{Computer Science Department\\
    \textit{William} \textit{\&} \textit{Mary}\\
    Williamsburg, USA\\
    achen08@wm.edu}
  \and
  \IEEEauthorblockN{Yanfu Yan}
  \IEEEauthorblockA{Computer Science Department\\
    \textit{William} \textit{\&} \textit{Mary}\\
    Williamsburg, USA\\
    yyan09@wm.edu}
  \and
  \IEEEauthorblockN{Denys Poshyvanyk}
  \IEEEauthorblockA{Computer Science Department\\
    \textit{William} \textit{\&} \textit{Mary}\\
    Williamsburg, USA\\
    dposhyvanyk@wm.edu}
}

\maketitle

\begin{abstract}
We introduce ACER, an AST-based call graph generator framework. ACER leverages tree-sitter to interface with any language. 
We opted to focus on generators that operate on abstract syntax trees (ASTs) due to their speed and simplicitly in certain scenarios; however, a fully quantified intermediate representation usually provides far better information at the cost of requiring compilation.
To evaluate our framework, we created two context-insensitive Java generators and compared them to existing open-source Java generators.

Code: \href{https://github.com/WM-SEMERU/ACER}{https://github.com/WM-SEMERU/ACER}

\end{abstract}


\section{Introduction}
\label{sec:intro}

Long-standing research endeavors in static call graph generation have made significant contributions to the field of program analysis over the years. Call graphs can be used in a wide variety  of areas: 
from critical tasks like compiler optimization \cite{10.5555/286076} and detecting security vulnerabilities \cite{10.1016/j.jnca.2015.11.017}, to useful applications such as code profiling \cite{10.1002/spe.562}, refactoring \cite{10.1145/2076021.2048078}, and navigation\cite{10.1109/icse.2013.6606621}. Formally, a call graph is defined as a directed graph $G=(V, E)$ wherein each vertex $v \in V$ represents a method, and an edge $e=(v_{1}, v_{2}) \in E$ signifies that the method $v_1$ invokes the method $v_2$ within its body. 

\lstset{
  language=Java,
  commentstyle=\color{green},
  keywordstyle=\color{blue},
  stringstyle=\color{red},
  basicstyle=\ttfamily,
}

\begin{figure}[!h]
\begin{lstlisting}
class Bar {
    void bar() {}
}
public class Foo {
    Bar b = new Bar();
    void method1(Bar b) {
      b.bar();
    }
}
\end{lstlisting}
\caption{For jargon demonstration purposes.}
\label{lst:jargon-demonstration}
\end{figure}
In call graph terminology, an invocation is known as a \textit{call site}. There exist two call sites in \autoref{lst:jargon-demonstration}: \texttt{new Bar()} and \texttt{b.bar()}. The main operation of call graph generators lies in call site resolution — the process of identifying the fully quantified method(s) a call site corresponds to. In Java, a method is considered fully quantified when its package, enclosing classes, name, and argument types are all taken into account.

A call site's left hand side (if it exists) is referred to as the \textit{receiver}. In \texttt{b.bar()}, the receiver is \texttt{b}.
\textit{Call containers} are nodes that lexically contain call sites. There are two call containers in the figure: \texttt{class Foo} and \texttt{void method1(Bar a)}.

Call graph generators vary mainly by the following three parameters: algorithm, source format, and scope of analysis. We list examples in \autoref{callgraph-categories}.

\begin{table}[ht]
\centering
\normalsize
\begin{tabular}{|l|p{5cm}|}
\hline
\textbf{Paramters} & \textbf{Examples} \\
\hline
Algorithm & NR, CHA, SCHA, RTA, k-CFA\\
\hline
Scope of Analysis & Application-only, Library-only, In-between, Full software \\
\hline
Source Format & Raw source, AST, IR\\
\hline
\end{tabular}
\caption{Generator parameters and examples.}
\label{callgraph-categories}
\end{table}

Generators first differ drastically by their algorithm. Grover et al. \cite{10.1145/506315.506316} established a taxonomy for call graph algorithms over twenty years ago, which Tip et al. \cite{10.1145/354222.353190} broadly categorized into three levels: 
\begin{enumerate}

\item Simple and few-passed, such as Name-based Resolution (NR) and Class Hierarchy Analysis (CHA). NR disregards receivers and resolves call sites using solely the method name. CHA, on the other hand, considers receivers and even their subtypes to support polymorphism.
\item Context-insensitive but incorporates some points-to analysis. Examples include Variable Type Analysis (VTA) and Andersen-style Pointer Analysis (APA) \cite{Andersen2005ProgramAA}.
\item Context-sensitive, exemplified by $k$-CFA ($k$-Control Flow Analysis), where $k$ means to utilize the last $k$ call container contexts.
\end{enumerate}

Call graph generators also vary in their scope of analysis, the part of the full software analyzed. By full software, we mean the application source plus its library dependencies.
There are four levels of scope of analysis (from narrow to broad):
\begin{enumerate}
    \item Application-only: Only includes edges between application methods.
    \item Library-only: While libraries by themselves could just be regarded as applications, Reif et al. introduced algorithms dedicated to libraries \cite{10.1145/2950290.2950312}. The key distinction is that libraries could be analyzed from the user's perspective and assume certain private methods to be closed off. This constructs a smaller set of entry points and thus a more precise graph.
    \item  In-between: This contains application-only edges and library methods invoked in application code \cite{10.1007/978-3-642-31057-7_30}. 
    \item Full-software: This contains in-between edges and includes library methods invoked in library code.
\end{enumerate}

Lastly, generators differ by their source format. \textit{Source-based} generators use the raw source files while \textit{AST-based} generators have access to the abstract syntax trees. The choice of algorithm is tied with the source format — source-based generators are likely to implement the simplest NR algorithm because extracting information from unstructured raw code is hard and unscalable. Examples of source-based generators include the internal utility within Doxygen \cite{doxygen} and this Perl-based multilingual generator \cite{koknat2023callGraph}. Examples of AST-based generators are the popular Python generators \cite{pyan} \cite{10.1109/icse43902.2021.00146}, which operate on the outputs of the \texttt{ast} and \texttt{symtable} modules.

Additionally, there exist \textit{IR-based} generators, which operate on some internal representation (IR) created during compilation. Typically, only statically-typed languages have IRs. IRs deliver information that neither raw sources nor ASTs directly provide — the most important data being fully-qualified functions. Since a call graph's vertices are just fully-resolved functions, the construction of a sound call graph from IR is simple (but not trivial due to polymorphism) because IR has already resolved the functions. AST-based generators, on the other hand, need to implement resolution logic.

Examples of IR include JVM bytecode \cite{lindholm2014java} and the fully-resolved syntax tree outputted by Clang-based compilers during semantic analysis \cite{clang_ast}. We consider Clang's rich tree to be IR and no longer just an AST. Tools that generate IR usually require the source to be compilable — a program's dependencies thus must be pulled in and built. Ideal and error-resilient tools like Roslyn \cite{Roslyn} make exceptions — they can generate the IR without building the dependencies because they resolve all resolvable entities and simply flag the unresolvable ones. On the other hand, most tools like Clang AST and GCC \cite{gcc} either throw or remove unresolvable entities.

As an aside, languages that can compile to JVM bytecode can get call graph generators ``for free" through bytecode analysis frameworks WALA \cite{WALA} and Soot \cite{Soot}. However, the implementation details matter greatly — JVM-hosted implementations of Groovy, Clojure, Python, and Ruby produce very unsound call graphs due to dynamic translation schemes \cite{10.1109/tse.2019.2956925}.


Although IRs contain more useful data than ASTs, we still build our call graph framework based on AST for the following reasons:
\begin{enumerate}
    \item Compile-free: Most IRs can only be generated if the source is compilable, a restriction that AST-based generators do not impose. Further, IR compilation may compute information irrelevant to generators. Consequently, AST-based generators theoretically perform faster.
    \item Generalizable: Most languages can be parsed to an AST while only some statically-typed languages have compilers that output rich, fully-resolved IRs.
    \item  Synergizes with the application-only scope: If the goal is to generate application-only edges, there's no need to consider the dependencies. AST-based generators complement this by operating directly on ASTs from application source. Conversely, IR-based generators must take into account the dependencies, as most IRs can only be produced when the entire software is compilable.

\end{enumerate}

We have outlined the key generator parameters and provided reasons for using AST-based generators. We now turn to the focus of the paper — how to construct AST-based generators.

The first piece of our approach is the powerful tree-sitter \cite{tree-sitter} library. Tree-sitter is both a parser generator tool and an incremental parsing library. It provides parsers for all popular languages and a common interface to interact with the parsers and their concrete syntax trees outputs. Though the common interface is written in C, tree-sitter provides additional language-bindings — we use the Python binding.

The second piece of our approach consists of language-agnostic components and utilities. These abstract the shared logic common to generators across various algorithms, scopes, and languages. These components are detailed in \autoref{tool-internals}. Together, these two pieces form the backbone of our framework, ACER, designed to simplify the development of call graph generators.

Here are our main contributions: 
\begin{enumerate}
    \item Introduced ACER, an \textbf{A}ST-based \textbf{C}allgraph G\textbf{E}nerator F\textbf{R}amework.
    \item Provided a taxonomy for generator design choices and document implementational challenges. 
    \item Developed and evaluated a single NR and a single SCHA (Simple CHA) AST-based Java generator using ACER.

\end{enumerate}

\section{Background and Related work}
\label{sec:Background and Related work}

In the last section, we introduced the main call graph parameters: algorithms, source formats, and scope of analysis. In addition to these main traits, there exist subtle details like reachability. In this section, we expand on all parameters and discuss the challenges in implementing them.

\subsection{Algorithm}
\label{subsec:Algorithm}
An advanced call graph algorithm builds upon a simpler one by either pre-caching supplementary structures or monitoring additional runtime data. High recall and soundness is easier to achieve while high precision demands significantly more computational power \cite{10.1145/354222.353190}.

As previously mentioned, the source format is tied intimately with the algorithm. The Java bytecode IR, if interpreted directly, encourages an algorithm that is more precise than NR but less sound than CHA. Surprisingly, four out of the six most popular Java generators use this direct interpretation and thus do not fully support CHA. We illustrate the difficulty of fully supporting CHA in \autoref{lst:polymorphism}.

\lstset{
  language=Java,
  commentstyle=\color{green},
  keywordstyle=\color{blue},
  stringstyle=\color{red},
  basicstyle=\ttfamily,
}

\begin{figure}[!h]
\begin{lstlisting}
class A { void method() {}; }
class B extends A {}
class C extends B {}

class Bar {
    void foo(A a) { // Could be A, B, C
      a.method();
    }
}
\end{lstlisting}
\caption{Liskov's substitution principle \cite{10.1145/197320.197383} in play.}
\label{lst:polymorphism}
\end{figure}
The call site \texttt{a.method()}, in JVM bytecode, is represented as \texttt{1: invokevirtual \#2 // Method A.method:()V}. A simple, direct generator deduces the edge from this information alone, thereby including only the edge from \texttt{Bar.foo} $\rightarrow$ \texttt{A.method}. However, polymorphism also allows objects of type \texttt{B} and \texttt{C} to be passed into \texttt{foo}. Thus, sound analysis must include the additional edges \texttt{Bar.foo} $\rightarrow$ \texttt{B.method} and \texttt{Bar.foo} $\rightarrow$ \texttt{C.method}.

To support CHA, the generator must pre-cache a class hierarchy. After the type of a receiver is calculated, the generator considers the receiver to inhabit any subtype of the type calculated.
To support VTA and k-CFA, the generator also need to consider intraprocedural assignments and data flow.

\subsection{Handling Language Features}

Practical call graph generators abiding by the same algorithm and scope of analysis can still differ due to the attention to details \cite{10.5220/0007929201170128}. Some Java generators might not regard static initializer blocks as call containers, or disregard default interface methods. Conversely, some Python generators may not handle higher-order functions and iterators. In the last subsection, we also touched on the importance of handling polymorphism, which is addressed by building and using a class hierarchy.



\subsection{Scope of Analysis}
\label{scope-of-analysis}
Earlier, we mentioned that AST-based generators synergize with application-only scope. However, for broader scopes starting from in-between, IRs are preferred because most IRs compile the libraries by default.

As a side note, what Karim et al. \cite{10.1007/978-3-642-31057-7_30} regards as ``application-only'' is what we consider  as ``in-between". Their framework, \texttt{CGC}, parses just the structural information of the libraries to efficiently include edges from application to library and vice versa.

An additional parameter in the realm of scope is reachability. The most sound analysis considers all methods as potential entry points to handle potential multi-threading \cite{10.1145/2950290.2950312}. Precise generators may consider a smaller set of entry points (the singleton set of just the main method).

\subsection{Ambiguity Problems}
\label{ambiguity-problems-section}
Application-only, AST-based generators face ambiguity problems. Because IRs come from compilation, they have fully-resolved all entities. But, application-only AST-based generators can never fully resolve because they do not have access to the library sources. We illustrate this ambiguity in \autoref{lst:ambiguity-problem}.
\begin{figure}[!h]
\begin{lstlisting}
import java.util.*;

class Bar {
    int add(int a, int b) {...}
    int add(float a, float b) {...}
    void foo(List<?> l1, List<?> l2) {
      int sum = add(l1.size(), l2.size());
    }
}
\end{lstlisting}
\caption{Without library analysis, we do not know the type of \texttt{l1.size()} and \texttt{l2.size()}. 
So, we can at best return all methods named \texttt{add} with two arguments.}
\label{lst:ambiguity-problem}
\end{figure}
For a reasonable implementation, methods in the figure should only aim to be identified by their package, class, method names, and the number of arguments, which creates unnecessary edges.

In-between scoped and AST-based generators like PyCG do not face this problem. PyCG uses \texttt{importlib} to attempt to resolve library definitions.

\section{Tool Description and Internals}
\label{tool-internals}
\begin{figure}[h]
\centering
\includegraphics[width=1\columnwidth]{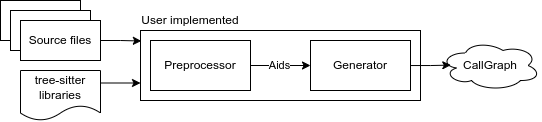}
\caption{Call graph Generation Flow. \textit{Preprocessor} and \textit{Generator} are user-implemented.}
\label{fig:callgraph-pipeline}
\end{figure}

In this section, we describe the two essential classes of ACER: \textit{Preprocessor} and \textit{Generator}. We also show how ACER can be used to implement a hypothetical, application-only, AST-based, and fully-typed CHA Java generator. Generators built using ACER are by default confined to ASTs for their source format and initiate their scope of analysis at the application-only level. The algorithm, reachability, and the treatment of language features can be freely explored.

Users create new generators by extending the \textit{Preprocessor} and the \textit{Generator} and implementing a handful of abstract methods. Internally, we make use of strict type hints and generics to guide implementation.

\subsection*{Preprocessor}
The sole purpose of the \textit{Preprocessor} is to cache lookup structures that the \textit{Generator} utilizes. In this section, we first describe the inputs and outputs of the \textit{Preprocessor}. Then, we give a high level description of our CHA generator's \textit{Preprocessor} implementation.

\textit{Preprocessor} first loads the tree-sitter libraries: The C runtime and the input, language-specific parser. \textit{Preprocessor} then parses the input source files into tree-sitter trees. These trees are composed of tree-sitter \texttt{Node}s, which contain information like node type, text, and pointers to parent and children.

Provided with the roots of these trees, the \textit{Preprocessor} then produces its outputs. These outputs are the desired lookup structures. The \textit{Preprocessor} class has two abstract methods, which mandates the constructions of the two following structures:

\begin{enumerate}
    \item \texttt{method\_dict}: A mapping between unique method keys to their method bodies (tree-sitter \texttt{Node} objects). Our Java CHA generator creates the \texttt{method\_dict} by first locating all nodes of types \texttt{method\_declaration} and \texttt{constructor\_declaration}. From these nodes, the generator navigates upwards to locate their enclosing class and package for full resolution.
    \item  \texttt{unique\_dict}: A cache between non-unique keys to unique keys. The \texttt{unique\_dict} exists due to the application-only ambiguity problem described in \autoref{ambiguity-problems-section}: Call sites can not always resolve their argument types and should instead be identified by their argument count, but this introduces plurality which we keep track of through \texttt{unique\_dict}.
\end{enumerate} 

While every \textit{Preprocessor} must cache \texttt{method\_dict} and \texttt{unique\_dict}, advanced algorithms require more structures to be cached. Our Java CHA \textit{Preprocessor} builds two additional structures: 

\begin{enumerate}
    \item \texttt{package\_importables}: Contains the mapping from package to its exports (\eg fully quantified classes, interfaces, and enums). This aids full type resolution.
    \item  \texttt{class\_cache}: Contains the mapping from each class to its fields, subclasses, and method signatures. The cached fields and method signatures help resolve complex receivers. The cached subclasses mappings effectively form a class hierarchy, which the generator uses to handle polymorphism.
\end{enumerate} 
 
Both of these structures are created in a similar manner to how \texttt{method\_dict} was created — a parallelizable operation that runs on each file. All of these structures are then passed to the \texttt{Generator}.

\subsection*{Generator}
The \textit{Generator} creates the call graph from the specified source files. Similar to the previous section, we first describe the I/O behavior of the \textit{Generator}. Then, we discuss the internal flow of the \textit{Generator} before concluding with how we can implement the hypothetical CHA \textit{Generator}.

The inputs to the \textit{Generator} are the source files, entry points, and the \textit{Preprocessor} results. By default, the entry points are all methods (the keys of \texttt{method\_dict}). Users can narrow down these entry points; for example, they can filter out methods whose names are not ``main''.

We now turn to explaining the generation flow, summarized by the pseudocode in \autoref{alg:generator-flow}:

\begin{algorithm}[h]
    \SetAlgoLined 
    \KwIn{Preprocess Results $P$, Entry Points $EP$} 
    \KwData{$AnalysisDeque$}
    \KwOut{Call graph $G(V,E)$ } 
    $initializeDeque(EP)$
    
    $visited = empty\_set()$
    
    \While{$AnalysisDeque.not\_empty()$}{ 
        \If{$call\_site.id \in visited$}{ 
            $continue$
        } 
        $visited.add(call\_site.id)$
        
        $next\_contexts = resolve(context, call\_site)$

        \ForEach{$context \in next\_contexts$ }{
$call\_container$ = $context.call\_container$

$call\_sites = seek\_call\_sites(call\_containers)$ 
           
            \ForEach{$call\_site \in call\_sites$} {
                 $AnalysisDeque.add((context, call\_site))$
            }
        
        }

    }

    \Return{$G = (V, E)$} 
    \caption{Generation Flow}
    \label{alg:generator-flow}
\end{algorithm} 

We clarify a few aspects. First, similar to \textit{Preprocessor}, \textit{Generator} is also an abstract class. It requires two abstract methods to be implemented: \texttt{seek\_call\_sites} and \texttt{resolve}. \texttt{seek\_call\_sites} finds new call sites to be added to the analysis from call containers. \texttt{resolve} determines the fully quantified function(s) that a call site corresponds to.

The generation algorithm keeps the following two runtime structures:
\begin{enumerate}
    \item \texttt{AnalysisDeque}: An explicit deque is used to keep track of the call sites to resolve. By default, the algorithm automatically adds call sites contained by methods. This explicitness, however, allows users to add call sites contained by non-method call containers. 
    \item \texttt{context}: Caches data-flow and points-to analysis results required by VTA and k-CFA. 
\end{enumerate}


Our CHA generator must implement \texttt{seek\_call\_sites} and \texttt{resolve}. Our generator seeks call sites by returning all instances of \texttt{method\_invocation} and \texttt{object\_creation\_expression} nodes within call containers. 

Implementing \texttt{resolve} requires more sophistication. We show just the logic to \texttt{resolve} a \texttt{method\_invocation} node, which looks like \texttt{receiver.method(...)}. The receiver may be of any expression. In JDK8, we found that implicit/explicit \texttt{this} and identifiers make up 88\% of the receivers while \texttt{field\_access} and \texttt{method\_invocation} nodes only made up 8\%. A call site whose receiver is of type \texttt{field\_access} will look like \texttt{a.b.c}, where the receiver is \texttt{a.b}.; a call site whose receiver is a \text{method\_invocation} node looks like \texttt{a.b().c}, where the receiver is \texttt{a.b()}.

In all of these scenarios, we must first resolve the full type of the identifier \texttt{a}. We implement this by walking upwards from the identifier to find where it was introduced. We scan the declaration statements, method arguments, and class fields. However, we can only expect to retrieve the shorthand type from these declarations because Java supports type aliasing. To resolve the full type, we build a list of possible alias to full type mappings by analyzing the import statements and utilizing \texttt{package\_importables}. Then, we determine the full type by querying the aliased type against the built list. If the query yields no results, the aliased type is probably defined by the libraries, which the application-only scope ignores. Once the receiver's full type is resolved, we address polymorphism by identifying the type's subclasses through a query to \texttt{class\_cache}. 



With the full type of the identifier \texttt{a} determined, call sites like \texttt{a.b()} can be easily resolved. To resolve scenarios like \texttt{a.b.c()} and \texttt{a.b().c}, we just have to utilize the lookup structures generated by the \textit{Preprocessor} a bit more.

We have covered the high level overview of building a Java CHA generator with ACER. Evidently, much of the user implemented logic deals with type resolution, which comes with freely with IR.

\begin{table*}[ht]
\centering
\begin{tabular}{|l|r|r|r|r|r|r|r|r|r|r|}
\hline
& {Soot} & {OSA} & {SPOON} & {JCG} & {WALA} & {JDT} & {NR} & {NR-ALL} & {SCHA} & {SCHA-ALL} \\ \hline
Soot & 34574 & 44.6\% & 41.3\% & 42.4\% & 16.4\% & 44.0\% & 59.8\% & 69.5\% & 1.7\% & 25.5\% \\ \hline
OSA & 45.8\% & 33685 & 45.5\% & 51.1\% & 12.8\% & 56.5\% & 56.4\% & 65.0\% & 1.4\% & 25.5\% \\ \hline
SPOON & 87.7\% & 94.3\% & 16263 & 82.7\% & 27.0\% & 92.2\% & 64.6\% & 76.6\% & 2.5\% & 40.1\% \\ \hline
JCG & 64.3\% & 75.4\% & 58.9\% & 22834 & 17.8\% & 74.4\% & 53.8\% & 62.1\% & 2.1\% & 35.7\% \\ \hline
WALA & 100.0\% & 75.8\% & 77.4\% & 71.5\% & 5668 & 71.3\% & 66.2\% & 73.3\% & 6.1\% & 41.1\% \\ \hline
JDT & 78.1\% & 97.8\% & 77.1\% & 87.3\% & 20.8\% & 19463 & 70.5\% & 81.9\% & 2.1\% & 33.6\% \\ \hline
NR & 5.4\% & 4.9\% & 2.7\% & 3.2\% & 1.0\% & 3.6\% & 384315 & 100.0\% & 0.2\% & 3.5\% \\ \hline
NR-ALL & 5.5\% & 5.0\% & 2.9\% & 3.2\% & 1.0\% & 3.7\% & 88.0\% & 436726 & 0.2\% & 3.4\% \\ \hline
SCHA & 18.0\% & 14.1\% & 12.5\% & 14.6\% & 10.7\% & 12.5\% & 23.6\% & 24.4\% & 3242 & 100.0\% \\ \hline
SCHA-ALL & 14.5\% & 14.2\% & 10.8\% & 13.4\% & 3.8\% & 10.8\% & 21.9\% & 24.7\% & 5.3\% & 60677 \\ \hline
\end{tabular}
\caption{Common calls of the ArgoUML project between 6 existing generators and 4 of our generators. } 
\label{edges-output}
\end{table*}

\section {Examples and Evaluation}
\label{examples-evaluations}

Under ACER, we implemented a NR generator and a SCHA generator. We elaborate on our SCHA algorithm. 
	 
\subsection*{SCHA (Simple Class Hierarchy Analysis)}
Our SCHA generator is simple in comparison to the hypothetical CHA generator of \autoref{tool-internals}. First, our generator uses aliases, so collisions will occur if two methods belong to classes of the same alias and have the same name and number of arguments.
Second, only call sites with a simple identifier receiver like \texttt{a.b()} and those with no callers are considered. Call sites with complex receivers are ignored (\eg \texttt{(builder.make\_class()).method()}.



\subsection*{Evaluation}
Our evaluation goals revolve around measuring framework-centric metrics, \eg speed and lines-of-code. However, to measure our speed performances against existing tools, we cannot directly compare the call graph generation times. Speed depends on multiple factors, the most important being the preciseness of the algorithm. Given that the algorithms of the evaluated generators are quite different, we first present edge comparisons across all tools. This establishes a baseline level of precision that serves to validate the more important measurements.

We borrow the edge results of six open-source Java generators from Jász et al.  \cite{10.5220/0007929201170128}. They evaluated existing tools on ArgoUML, a sufficiently large, UML diagramming application written in Java. They thoughtfully performed full-software, in-between, and application-only analysis. Because our generators are application-only, we care only for the application-only findings, which are shown in \autoref{edges-output}. 

The cells along the diagonal represents the number of edges the corresponding generator created. The other cells represents the ratio of the edges discovered by the row generator that were also discovered by the column generator. For example, 44.6\% of Soot's edges were also discovered by OSA. The last four rows and columns show the edge performances of our tool. The ``ALL'' suffix signifies that a generator considers the entry points to be the set of all methods.

As expected, the NR generators have high recall but low precision — they contain, on
average, 52.1\% of the edges that other tools generated. But other tools, on average, only contain 3\% of the edges NR generated. Conversely, the SCHA generators have higher precision but lower recall.

Generating more edges seems to correlate with a faster and sounder algorithm while generating less edges seem to imply a slower and more precise algorithm. This observation is indeed true for our sound NR tool, which had the most edges, and WALA, which ran an approximation of the APA algorithm. In reality, this is not always the case. The difference in choosing which language features to handle drastically change the edges. In the case the SCHA tool, it generated the least edges due to its simplicity (\eg only resolving call sites with identifier receivers) and not its algorithm rigor (which falls behind WALA's APA and Soot's CHA).

We admit that further detailed analysis on why the edges differ should be conducted. But, with this anchor set, we now turn to discuss speed performances.

\begin{table}[ht]
\centering
\normalsize
\begin{tabular}{|l|l|l|}
\hline
\textbf{Generator} & \textbf{Preprocessing Time} & \textbf{Generation Time}\\
\hline
NR  & 00:32 & 01:23  \\
\hline
SCHA &  00:59 &02:29   \\
\hline
JCG &  20:00&  00:07  \\
\hline
WALA &  20:00 &01:50   \\
\hline
\end{tabular}
\caption{Generator speeds in mm:ss}
\label{Generator-speed}
\end{table}


We evaluated the time it took for JCG, WALA (in CHA mode), NR-all, and SCHA-all to operate on JDK8's jar, on a machine with a Intel® Core™ i7-11800H processor. We zoomed in on JCG because it most closely resembles NR-ALL in terms of straightforwardness and WALA (in CHA mode instead of APA) because it most closely resembles SCHA. At last, we consider all methods as entry points because JDK8 is a library and does not have main methods. 

Per the table, other tools outperform our tools in terms of generation speed. But, this can be reasonably attributed to the fact that IR has already fully quantified all entities. For a fair comparison, we must consider the preprocessing time it took to build the heavy JDK8 jar, which takes 20 minutes. For our tools, we attribute preprocessing time to the \textit{Preprocessor} and generation time to the \textit{Generator}. With the preprocessing time considered, our tools perform much faster.

At last, we note that the framework itself, including utilities, contains 800 lines of code in Python. The NR generator was written in only 100 lines and the SCHA took 300 lines. The additional lines in SCHA went towards generating class hierarchy and resolving the full type of identifiers. We actually wrote two incomplete CHA generators for Java and Python that attempted to resolve all expression types. They came out to around ~3K lines — most lines going towards type resolution. Thus, we plan to research further in designing language-agnostic abstractions for assisting entity resolution. As mentioned before, the ideal solution is a source format that is fully quantified wherever possible, resilient to errors, and does not require full compilation and dependency building.

\section{Conclusion and Future Work}
\label{future-work}
In this paper, we presented ACER, an AST-based call graph generator framework. First, we introduced the parameters of call graph generators  and highlighted implementation challenges. Then, we discussed the process in building a hypothetical CHA generator under ACER. At last, we evaluated our Java NR and SCHA generators against existing, IR-based Java generators. 

Here is a list of items we plan to address in our future work:
\begin{enumerate}
	\item Conduct an in-depth analysis on edge outputs of different generators on JDK8. Build a table similar to \autoref{edges-output} and attribute edges variations to specific design choices.
 \item Build abstractions to assist name resolution and full quantification.
 \item Build generators for other languages to demonstrate the multilinguality of the framework.
\end{enumerate}

\section{Acknowledgements}
\label{Ack} We would like to thank the CSci 435 students Vinny Allegra, Daniel Lee, Aamir Mohammed, and Chas Rinne from Fall 2022 for contributing to the initial version of the project. This research has been supported in part by the NSF CCF-2311469, CNS-2132281, CCF-2007246, and CCF-1955853. We also acknowledge support from Cisco Systems. Any opinions, findings, and conclusions expressed herein are the authors’ and do not necessarily reflect those of the sponsors. 

\bibliography{./bib.bib}

\end{document}